\documentclass[12pt,letterpaper]{article}

\usepackage{amsmath}
\usepackage{amssymb}
\usepackage{color}
\usepackage{graphicx}

\numberwithin{equation}{section}

\def\SU{\mathrm{SU}}
\def\SL{\mathrm{SL}}

\def\U{\mathrm{U}}

\def\cA{\mathcal{A}}

\def\cH{\mathcal{H}}

\def\cN{\mathcal{N}}

\def\cS{\mathcal{S}}

\def\bR{\mathbb{R}}

\def\diag{\mathrm{diag}}
\def\ket#1{|#1\rangle}

\def\vev#1{\langle#1\rangle}

\let\sym\relax

\begin{document}

\begin{titlepage}
\begin{flushright}
\normalsize
PUPT-2376

\medskip

May, 2011
\end{flushright}
\vfil

\bigskip

\begin{center}
\LARGE 3d Partition Function as Overlap of Wavefunctions
\end{center}

\vfil
\medskip

\begin{center}
\def\thefootnote{\fnsymbol{footnote}}

Tatsuma Nishioka$^\heartsuit$,  Yuji Tachikawa$^\clubsuit$\footnote[1]{on leave from IPMU, the University of Tokyo}, and Masahito Yamazaki$^\spadesuit$

\end{center}

\begin{flushleft}\small
$^\heartsuit$ Department of Physics, Princeton University, Princeton, NJ 08544, USA

$^\clubsuit$ School of Natural Sciences, Institute for Advanced Study, Princeton, NJ 08540, USA

$^\spadesuit$
Princeton Center for Theoretical Science, Princeton University, Princeton, NJ 08544, USA
\end{flushleft}

\begin{figure}[h]
\begin{center}
	\scalebox{0.4}{\input{Relation_simp.pstex_t}}
\end{center}
\end{figure}

\end{titlepage}

\thispagestyle{empty}

\setcounter{tocdepth}{1}

\begin{center}
{\bfseries Abstract}
\end{center}

\bigskip

We compute the partition function on $S^3$ of 3d $\cN=4$ theories which
 arise as the low-energy limit of 4d $\cN=4$ super Yang-Mills theory on
 a segment or  on a junction, and propose its 1d interpretation.
We show that the partition function can be written as an overlap of
 wavefunctions determined by the boundary conditions. 
We  also comment on the connection of our results with the 4d
 superconformal index and the 2d $q$-deformed Yang-Mills theory.

\vspace{3cm}

\tableofcontents

\newpage
\setcounter{page}{1}
\section{Introduction}

The aim of this note is to study the partition function on $S^3$ of a class of 3d $\cN=4$ theories which arise as the 4d $\cN=4$ theory with gauge group $G=\SU(N)$ put on a segment or a junction.
We will see that the resulting partition function can be understood in terms of an overlap of wavefunctions, associated to the boundary conditions of the segment or the junction.

\begin{figure}[h]
\begin{center}\[
\vcenter{\hbox{\scalebox{0.45}{\input{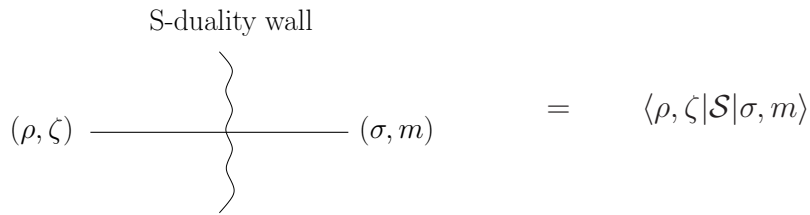}}}} \qquad\qquad \qquad \vcenter{ \hbox{$= \qquad \vev{\rho,\zeta|\cS|\sigma,m}$}}
\]
\end{center}
\caption{A graph with two boundary conditions $\rho$ and $\sigma$. The wavy line represents an S-duality wall.
}
\label{fig:S_duality_wall2}
\end{figure}

The first class of theories we discuss is the $T^\sigma_\rho[\SU(N)]$ theory introduced in \cite{Gaiotto:2008ak}, where $\rho$ and $\sigma$ are partitions of $N$. This arises as the low energy limit of 4d $\cN=4$ $\SU(N)$ Yang-Mills on a segment with two boundary conditions labeled by $\rho$ and $\sigma$, with an S-duality wall between them (see Fig.~\ref{fig:S_duality_wall2}). 
This theory has mass deformations $\zeta$ and $m$ associated to $\rho$ and $\sigma$, respectively.
We evaluate the partition function of this theory using the matrix model reduction \cite{Kapustin:2009kz} and will find that it has a universal form \begin{equation}
Z[T^\sigma_\rho[\SU(N)]]=\vev{\rho,\zeta|\cS|\sigma,m} \ . \label{eq:r}
\end{equation} 
Here, the kets $\ket{\rho,\zeta}$, $\ket{\sigma,m}$ are states in the Hilbert space $\cH$, the space of wavefunctions 
defined on the Weyl chamber of $\SU(N)$  Lie algebra, and $\cS$ is essentially the Fourier transformation on it. 
This result has the following natural interpretation.
We have the 4d theory on $S^3$ times an interval. When the size of $S^3$ is small, 
the system can be regarded as a 1d theory on the segment,
with initial and final states determined by the boundary conditions.
The segment represents a time evolution and the S-duality wall inserts a corresponding operator $\cS$ to the theory.
By taking the  limit where the segment is also short, the time evolution drops out, and we have an overlap of wavefunctions \eqref{eq:r}.

\begin{figure}[h]
\begin{center}\[
\vcenter{\hbox{\scalebox{0.45}{\input{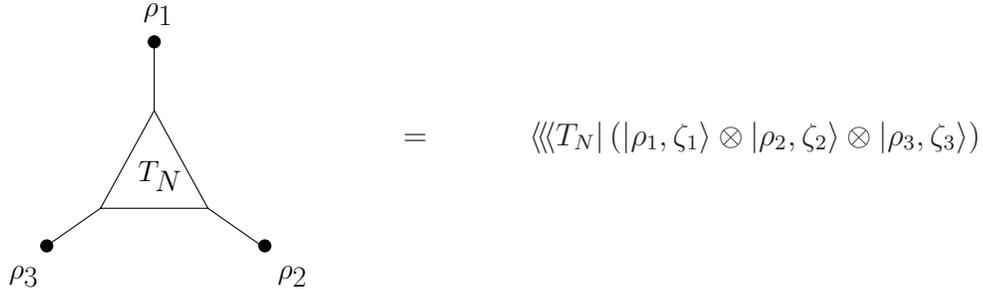}}}}
\qquad\qquad\vcenter{ \hbox{$= \qquad\quad \langle\!\langle \!\langle  T_N|\,
 (\ket{\rho_1,\zeta_1}\otimes \ket{\rho_2,\zeta_2}\otimes
 \ket{\rho_3,\zeta_3})$}}
\]
\end{center}
\caption{A trivalent graph with three boundary conditions $\rho_1$, $\rho_2$ and $\rho_3$. 
$\ket{T_N}\!\rangle\!\rangle$ acts as a boundary condition at the center of the trivalent vertex.}
\label{fig:trivalent}
\end{figure}

The second class of the theories we discuss is the 3d version of the $T^{\rho_1,\rho_2,\rho_3}_N$ theory introduced in \cite{Gaiotto:2009we}.
The 4d version of this theory is generically non-Lagrangian, and 
the expression of its superconformal index is recently conjectured in \cite{Gadde:2011ik} via its relation to the $q$-deformed two-dimensional Yang-Mills.
When  it is put on $S^1$ and the low-energy limit is taken, this theory has a mirror quiver
description \cite{Benini:2010uu} and the partition function can be  calculated using the partition function of $T^\sigma_\rho[\SU(N)]$. 
We find that the partition function is given by \begin{equation}
Z[T_N^{\rho_1,\rho_2,\rho_3}]=\langle\!\langle \!\langle  T_N|\,
 (\ket{\rho_1,\zeta_1}\otimes \ket{\rho_2,\zeta_2}\otimes
 \ket{\rho_3,\zeta_3})\label{eq:foo} \ ,
\end{equation}
where $\ket{T_N}\!\rangle\!\rangle$ is a  state defined in $\cH^3$ by  the $T_N=T_N^{[1,\ldots,1],[1,\ldots,1],[1,\ldots,1]}$ theory. See Fig.~\ref{fig:trivalent}.

We will see that this form \eqref{eq:foo} naturally arises as the small-radius limit of 
the superconformal index of the 4d  $T^{\rho_1,\rho_2,\rho_3}_N$ theory proposed in \cite{Gadde:2011ik}, following the procedure of \cite{Dolan:2011rp,Gadde:2011ia,Imamura:2011uw}. One essential ingredient in their proposal was the character $\chi_{R_\lambda}(U)$ of a group element $U$ in the representation $R$ whose highest weight is $\lambda$. The limit is taken by first setting  \begin{align}
2\pi m \sim -\beta\lambda \ , \quad 
U \sim \diag(e^{-i \beta \zeta_1},e^{- i\beta \zeta_2},\ldots,e^{- i\beta \zeta_N}) \ ,
\end{align} and letting $\beta\to0$, keeping $\zeta$ and $m$ fixed.
We will see that, roughly speaking, \begin{equation}
 \chi_R(U) \to \vev{\zeta|\cS|m} \ ,
\end{equation}  in the small $\beta$ limit.
Thus, two distinct mathematical objects, the representation $R$ and the group
element $U$ both become the Lie-algebra-valued objects $\zeta$ and $m$,
and are completely on the same footing in 3d.

The rest of the paper is organized as follows. 
After briefly reviewing the 4d $\cN =4$
theory on a segment in Sec.~\ref{subsec:segment},
we compute in Sec.~\ref{ssec:TSUN} and \ref{ssec:Trhosigma} the
partition function of the $T_\rho^\sigma[\SU(N)]$ 
theory by localization \cite{Kapustin:2009kz}, 
and find a succinct expression with manifest mirror symmetry.
In Sec.~\ref{ssec:1d}, we give an interpretation of the partition function as an overlap of wavefunctions
of a 1d theory on the segment. 
In Sec.~\ref{sec:comparison}, we show that the partition function of the 3d theory $T_N^{\rho_1, \rho_2, \rho_3}$
arises as a limit of the superconformal index of the 4d theory
$T_N^{\rho_1, \rho_2, \rho_3}$. We also
comment on the relation of our 1d system with the $q$-deformed 2d Yang-Mills
theory. 
We conclude with a short discussion in Sec.~\ref{sec:conclusions}.
We have a few appendices: in Appendix A, B, C and D, we provide further details of calculations, checks  and proofs. In Appendix \ref{sec:E}, we show the $E_n$  invariance of the partition function of the mirror quiver of the 3d $E_n$ theories.

\paragraph{Note:} When the authors barely started preparing the manuscript, the paper \cite{Benvenuti:2011ga} appeared, which has a substantial overlap of wavefunctions with this manuscript.  The paper \cite{Gulotta:2011si} also has a small overlap.

\section{$T^\sigma_\rho[\SU(N)]$}
\subsection{4d Super Yang-Mills on a Segment}\label{subsec:segment}
Let us put  4d $\cN=4$ theory with gauge group $G=\SU(N)$ on a segment, with half-BPS boundary conditions at the ends.
To describe the boundary conditions, we split six scalar fields $\Phi_{1,\ldots,6}$ of the theory into $\vec X=X_{1,2,3}$ and $\vec Y=Y_{1,2,3}$. From the viewpoint of the 3d theory, $\vec X$ is in the vector multiplet and $\vec Y$ is in the hypermultiplet.
Let us denote by $y$ the distance to the boundary.
A class of boundary conditions which preserves a fixed half of 32 supercharges is described by an embedding $\rho: \SU(2)\to \SU(N)$,
which controls the Nahm pole of $\vec X$ close to $y=0$:
\begin{equation}
X_i \sim {\rho(t_i)}/{y} \;,\label{eq:nahmpole}
\end{equation}
where $t_i$ $(i=1,2,3)$ are three generators of $\SU(2)$.
Note that $\rho$ is specified by a partition $N=n_1+n_2+\cdots+n_k$,
which we denote by $\rho=[n_1,\ldots,n_k]$. We order $n_i$ so that $n_i \ge n_{i+1}$.
Then \begin{equation}
\rho(t_3)=i\ \diag(\underbrace{\frac{n_1-1}2,\ldots,\frac{1-n_1}2}_{n_1},
\ldots,
\underbrace{\frac{n_k-1}2,\ldots,\frac{1-n_k}2}_{n_k}
) \ . \label{eq:rhot}
\end{equation}
Let $w_k$ the number of times the integer $k$ appears in the partition $\rho$. 
Then the subgroup of $\SU(N)$ commuting with $\rho(\SU(N))$ is given by \begin{equation}
G^\rho=S[\U(w_1)\times \U(w_2)\times \cdots \U(w_N)] \subset \SU(N) \ ,
\end{equation}
and it remains as the flavor symmetry acting on the boundary. 
We denote this boundary condition by $\rho_X$. We can similarly consider the boundary condition $\rho_Y$.
These boundary conditions preserve dilatation centered at the boundary.
One can modify the boundary condition  \eqref{eq:nahmpole} 
so that one has
\begin{equation}
X_i \sim {\rho(t_i)}/y + \delta_{i,3} m \ ,
\end{equation} 
where $m$ is in the Cartan of the Lie algebra of $G^\rho$: \begin{equation}
m=i\ \diag(\underbrace{m_1,\ldots,m_1}_{n_1},\ldots,
\underbrace{m_k,\ldots,m_k}_{n_k}) \ .
\end{equation}
When $\rho=[1,\ldots,1]$, there is no Nahm pole. 
We refer to this boundary condition  as the Dirichlet boundary condition.

Let us consider a domain wall, which we call the S-duality wall, 
across which the S-duality of $\cN=4$ super Yang-Mills is taken.
We put the 4d $\cN=4$ theory on a segment, with $\sigma_X$ on the left and $\rho_Y$ on the right, and the S-duality wall in between.
The low-energy limit is called $T_\rho^\sigma[\SU(N)]$ \cite{Gaiotto:2008ak}.
Recalling that $\vec X$ and $\vec Y$ are in the vector and the hypermultiplet, respectively, 
we find that $T^\rho_\sigma[\SU(N)]$ is the 3d mirror of $T_\rho^\sigma[\SU(N)]$.
For convenience, when $\rho$ or $\sigma$ is equal to $[1,\ldots,1]$, we often drop it from the notation. For example, $T_\rho[\SU(N)]$ is an abbreviation for $T^{[1,\ldots,1]}_\rho[\SU(N)]$.

\begin{figure}[h]
\begin{center}
\scalebox{0.45}{\input{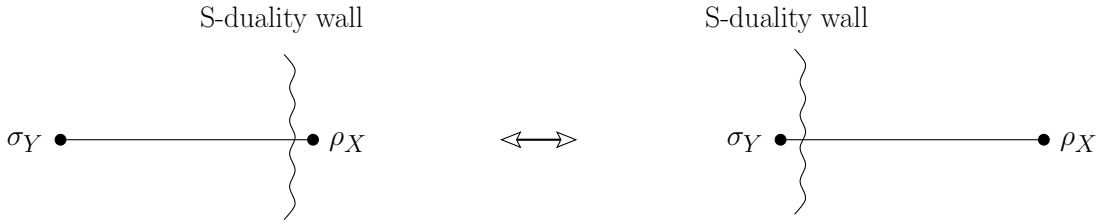}}
\end{center}
\caption{
Mirror symmetry between $T_\rho^\sigma$ and  $T_\sigma^\rho$ theories.  The low-energy theory is called $T_\rho^\sigma$
when the S-duality wall is close to the right 
boundary $\rho_X$, while it is called $T_\sigma^\rho$ when the wall is close to the left boundary $\sigma_Y$.}
\label{fig:S_duality_wall}
\end{figure}

4d $\cN=4$ $\SU(N)$ super Yang-Mills arises as the low-energy limit of the theory on $N$ D3-branes. Then, $\rho_X$ for $\rho=[n_1,\ldots,n_k]$ 
is realized by putting $k$ D5-branes extending along $\vec X$,
so that $n_i$ D3-branes end on the $i$-th D5-brane. 

\begin{figure}[h]
\begin{center}
\scalebox{0.45}{\input{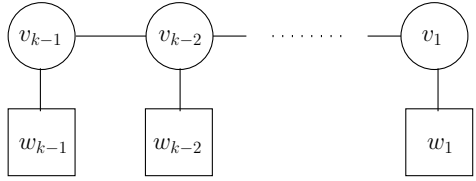}}
\end{center}
\caption{
The quiver for $T^\sigma_\rho(\SU(N))$ theory. A circle represents $\U(v_i)$ gauge groups,
a box represents $\U(w_i)$ flavor symmetry, and the line connecting two unitary groups 
represents hypermultiplets in the bifundamental representation.
\label{fig:generalquiver}}
\end{figure}

This allows us to find a linear quiver realization of $T_\rho[G]$:
we move the S-duality wall to the boundary specified by $\rho=[n_1,\ldots,n_k]$, 
turning D5-branes to   NS5-branes.
Rearranging them so that the low-energy gauge group can be straightforwardly read off,
we find that the low-energy limit is given by the quiver gauge theory shown in Fig.~\ref{fig:generalquiver}
with gauge groups $\U(v_i)$ ($i=1,\ldots,k-1$) where \begin{equation}
v_{k-1}=n_k \ ,\qquad 
v_i=v_{i+1}+n_{i+1} \qquad  \text{for}~ i=1, \dots, k-1 \ .
\end{equation}
with $N$ additional flavors for $\U(v_1)$.

\begin{figure}[h]
\begin{center}
\[
 \vcenter{\includegraphics[width=14cm]{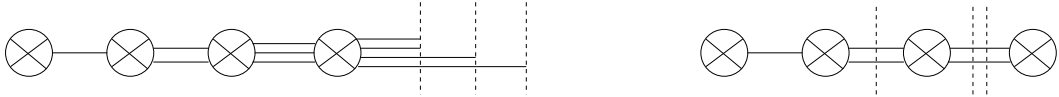}}
\]
\end{center}
\caption{
An example of brane construction of $T_\rho^\sigma[\SU(N)]$ theory with $\rho=[1,1,1,1]$ and $\sigma=[2,1,1]$. 
The circle ``$\otimes$" stands for an NS5-brane, and the dotted and solid lines denote D5- and D3-branes, respectively. When 5-branes are rearranged,  some of D3-branes are   annihilated due to the Hanany-Witten effect \cite{Hanany:1996ie}. 
}
\label{fig:brane}
\end{figure}

The quiver for $T^\rho_\sigma(\SU(N))$  can also be found by rearranging the NS5- and D5-branes.
See Fig.~\ref{fig:brane} for an example.
The result is again of the form given in Fig.~\ref{fig:generalquiver}.
It has gauge groups $\U(v_i)$  with $w_i$ additional flavors for $i=1,\ldots,k-1$.
Here,  $v_i$ and $w_i$ are related 
to $\rho=(n_i)$ and $\sigma=(\nu_i)$  as follows: \begin{equation}
\sigma=[\underbrace{k-1,\ldots,k-1}_{w_{k-1}},\ldots,\underbrace{s,\ldots,s}_{w_s},\ldots] \ ,
\end{equation} and \begin{equation}
n_k=v_{k-1}, \quad 
n_{i}=n_{i+1}-2v_i+(w_i+v_{i+1}+v_{i-1}) \quad \text{for}~ i=1,\ldots,k-1 \ .
\end{equation}
See an example in Fig.~\ref{fig:brane}.
This rule was originally written down from the consideration of the Higgs branch in \cite{Nakajima}; it can also be derived from the brane rearrangement in \cite{Gaiotto:2008ak}.

The theory $T_\rho^\sigma[G]$ is known to be empty \cite{Nakajima,Gaiotto:2008ak} unless $\sigma^T \ge \rho$,
where $\rho'\ge \rho$ is defined by the condition \begin{equation}
\sum_i^k n_i' \ge \sum_i^k n_i  \quad \text{for all} ~~  k \ ,
\end{equation}
and $\sigma^T$ denotes the transpose of $\sigma$. $\rho'\ge \rho$ is equivalent to $\rho^T \ge \rho'{}^T$.  This condition ensures $v_i>0$.

We can deform the boundary conditions by the mass deformations $m$, $\zeta$ for $\sigma$, $\rho$, respectively.  
In terms of the linear quiver, $m$ comes from the mass parameters of $w_i$ fundamentals,
and $\zeta$ comes from the  Fayet-Iliopoulos (FI) parameters $\alpha_i$ of $\U(v_i)$.

The partition function of this theory on $S^3$ can be exactly evaluated
by a matrix integral \cite{Kapustin:2009kz}, composed from the following ingredients:
\begin{itemize}
\item  For each gauge group $\U(v_i)$, we have integration variables $\sigma_i=(\sigma_{i,1},\ldots,\sigma_{i,v_i})$  with the measure \begin{equation}
\frac{1}{N!}\int \! d\sigma_i\, \Delta^2(\sigma_i) \ ,
\end{equation} where \begin{equation}
\Delta(\sigma)=\prod_{a<b} \sinh \pi(\sigma_a-\sigma_b) \ .
\end{equation}
\item We have a factor $\exp({2\pi i \alpha_i \sum_a \sigma_{i,a} })$ coming from the FI term.
\item Each bifundamental of $\U(v_i)\times \U(v_{i+1})$ contributes a factor in the integrand \begin{equation}
\prod_{a,b}\frac{1}{\cosh \pi(\sigma_{i,a}-\sigma_{i+1,b})}{\atop \ .}
\end{equation}
\item Each fundamental of mass $m$ of $\U(v_i)$ contributes a factor  in the integrand\begin{equation}
\prod_a \frac{1}{\cosh \pi(\sigma_{i,a}-m)}{\atop \  .}
\end{equation}
\end{itemize}
We denote the resulting partition function by $Z[T^\sigma_\rho[\SU(N)]](\zeta,m)$.

\subsection{$T[\SU(N)]$}\label{ssec:TSUN}
	\begin{figure}[htbp]
	\begin{center}
	\scalebox{0.5}{\input{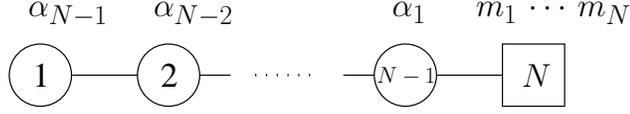}}
	\caption{A quiver diagram of the $T[\SU(N)]$ theory.}
	\label{fig:ZTSUN}
	\end{center}
	\end{figure}

We begin with the partition function of the theory $T[\SU(N)]$, whose quiver
diagram is shown in Fig.~\ref{fig:ZTSUN}.
We have mass parameters 
$m=(m_1,\ldots, m_N)$ and
FI parameters $\alpha=(\alpha_1, \ldots, \alpha_{N-1}) $.
We define $\zeta = (\zeta_1,\ldots, \zeta_N)$ via the relation
 $\alpha_i = \zeta_{i+1} - \zeta_{i} ~ (i=1,\ldots,N-1)$.
Both $m$ and $\zeta$ are subject to the constraints $\sum_i m_i = \sum_i \zeta_i = 0$
of the $\SU(N)$ group.

The partition function of this theory is given by a concise expression,
with manifest mirror symmetry \cite{Benvenuti:2011ga,Gulotta:2011si}:
\begin{align}
Z[T[\SU(N)]](\zeta,m)=\frac{1}{N!}\frac{\sum_{w\in W} (-1)^{w}
 e^{2\pi i m \cdot w
 (\zeta)}}{\Delta(m) \Delta(\zeta)} \ .
\label{ZTSUN}
\end{align}
Here, $\Delta(\zeta)$ is the $\sinh$ Vandermonde determinant
\begin{align}
\Delta(\zeta)=\prod_{i<j}\sinh\pi (\zeta_i-\zeta_j) \ ,
\end{align}
and $w$ is an element of the Weyl group $W$ of $\SU(N)$,
acting on the mass and FI parameters by permutation, {\it i.e.}, 
$m \cdot w (\zeta)=\sum_i m_i \, \zeta_{w(i)}$. 

It is natural to generalize the expression \eqref{ZTSUN} to an arbitrary
gauge group. The expression is manifestly invariant under the S-duality $G\leftrightarrow G^\vee$
\begin{align}
Z[T(G)](\zeta,m)=\frac{1}{|W|}\frac{\sum_{w\in W} (-1)^{l(w)}
 e^{2\pi i m \cdot w
 (\zeta)}}{\Delta_G(m) \Delta_{G^\vee}(\zeta)} \ ,
\label{ZTSUNG}
\end{align}
where $W$ is a Weyl group of the gauge group $G$, and $l(w)$ is the length
of the Weyl group element, and \begin{equation}
\Delta_G(m)=\prod_{\lambda\ :\ \text{roots of}\ G} \sinh(\pi\lambda\cdot m) \ .
\end{equation} 

\subsection{$T_{\rho}^{\sigma}[\SU(N)]$ } \label{ssec:Trhosigma}

Let us next describe the partition function of $T_{\rho}^{\sigma}[\SU(N)]$.
For $\rho=[n_1,\ldots,n_k]$, we have FI terms $\alpha_i$ assigned to the quiver. 
We then define $\zeta_{\rho}$ in the Cartan of $G$ as \begin{equation}
\zeta_{\rho} =\rho(t_3)+\diag( \underbrace{\zeta_{ 1},\ldots, \zeta_{ 1 }}_{n_1},\ldots,
\underbrace{ \zeta_{k},\ldots, \zeta_{ k}}_{n_k}  ) \ , \label{eq:zeta}
\end{equation} where \begin{equation}
\zeta_s=\sum_{t=0}^{s-1} \alpha_{t}\ , \qquad (\alpha_0 = 0)\ .
\end{equation}
See  Fig.~\ref{boxrule}a for an example.

For $\sigma=[\nu_1,\ldots,\nu_\kappa]$, we have mass terms $\tilde
m_{a,s}$ of the $s$-th fundamental hypermultiplet of $\U(v_a)$, where
$s=1,\ldots,w_a$. We then define $m$ in the Cartan of $G$ as 
\begin{equation}
m_{\sigma}=\sigma(t_3)+\diag( \underbrace{m_{\kappa},\ldots, m_{\kappa}}_{\nu_1}, \ldots,
\underbrace{m_{1},\ldots,m_{1}}_{\nu_\kappa}) \ , \label{eq:m}
\end{equation} where \begin{equation}
m=(\tilde m_{1,1},\ldots, \tilde m_{1,w_1},\ldots, 
\tilde m_{k-1,1},\ldots,\tilde m_{k-1,w_{k-1}})\ .
\end{equation}
See Fig.~\ref{boxrule}b for an example.

We then define a reduced Vandermonde determinant 
${\Delta}_\rho(\zeta)$ as a product of a Vandermonde determinant for 
variables associated with boxes in each row, {\it i.e.},
\begin{equation}
\Delta_\rho(\zeta)=\prod_{i}\prod_{a<b}\sinh \pi( (\zeta_{\rho}){}_{i,a}-(\zeta_{\rho}){}_{i,b}) \ ,
\end{equation}
where the indices $a,b$ run over boxes in a given row.

Then, we propose that the partition function of $T_\sigma^\rho[\SU(N)]$ is given by 
\begin{equation}
Z[T^{\sigma}_{\rho}[\SU(N)]](\zeta, m)=\frac{1}{N!}\frac{\sum_{w\in W} (-1)^w e^{2\pi
i m_{\sigma}\cdot w(\zeta_{\rho})} }{ \Delta_\sigma(m)\Delta_\rho(\zeta)
}\ . \label{eq:general}
\end{equation}
making  the mirror symmetry manifest.
A few examples confirming this proposal are presented in Appendix
\ref{sec:example}, and 
we present a proof of \eqref{eq:general} 
when $\sigma=[1,\ldots,1]$ in Appendix \ref{sec:proof}.
We provide another consistency check in Appendix \ref{sec:dominant}
by showing that our answer \eqref{eq:general} is non-zero only when $\sigma^T\ge \rho$,
which agrees with the fact that $T_{\rho}^{\sigma}$ is non-empty
only when $\sigma^T \ge \rho$, as we recalled in Sec. \ref{subsec:segment}.

\begin{figure}
\begin{align}
	\text{a)}&\quad \vcenter{\hbox{\scalebox{0.4}{\input{1-3-5-8.pstex_t}}}}
	\qquad\qquad\qquad\qquad\quad
	\vcenter{\hbox{\scalebox{0.4}{\input{YoungTab.pstex_t}}}} \nonumber\\[-2em]
	&\nonumber\\
	\text{b)}&\quad \vcenter{\hbox{\scalebox{0.35}{\input{mirror-1-3-5-8.pstex_t}}}}
	\qquad\qquad
	\vcenter{\hbox{\scalebox{0.4}{\input{YoungTab_mass.pstex_t}}}} \nonumber
\end{align}
	\caption{a) An example of the general rule  defining $\zeta$ in terms of  FI parameters associated with $T_\rho^{\sym} [\SU(N)]$ theory. 
	The reduced Vandermonde determinant becomes $\Delta_{\rho} (\zeta) = (\cosh\pi \alpha)^2 (\sinh\pi\beta)^2 \cosh\pi\gamma \, (\cosh\pi(\alpha+\beta))^2\cosh\pi(\beta+\gamma)\sinh\pi(\alpha+\beta+\gamma)$.	
	If the flavor symmetry is taken as $\U(8)$ symmetry and given an FI parameter $\delta$, all the parameters in the Young diagram will be shifted by $\delta$.
	b)  An example of the general rule for defining $m$ in terms of mass parameters associated with $T^\rho_{\sym} [\SU(N)]$ theory.
	This is mirror to the  theory above.}
	\label{boxrule}
\end{figure}

Note that the real parts of $\zeta$ and $m$ defined in \eqref{eq:zeta} and \eqref{eq:m} is in the Cartan of the flavor symmetry $G^\rho$ or $G^\sigma$, but the imaginary parts of $\rho(t^3)$  and $\sigma(t^3)$ are not.
In fact, the vector $\zeta$ is exactly of the form
that the state with momentum $\zeta$ in the Toda theory 
is a unitary, semi-degenerate state of type $\rho$  \cite{Kanno:2009ga,Drukker:2010vg}. 
This is natural from the point of view of \cite{Hosomichi:2010vh} where the $T^\sigma_\rho$ theory acts as a duality kernel of the Toda theory.

\subsection{1d Interpretation of the Partition Function}\label{ssec:1d}

The partition function of the $T[\SU(N)]$ theory \eqref{ZTSUN} 
is a Weyl-invariant kernel for the Fourier
 transformation\footnote{The role of the Fourier transformation in mirror symmetry is
discussed in \cite{Kapustin:1999ha,Kapustin:2010xq}.}.
Using the standard measure $\int dm\,\Delta(m)^2$ of the matrix model of \cite{Kapustin:2009kz}, 
one can prove the following identity:
 \begin{equation}
\int \! d\zeta \, \Delta^2(\zeta)\, Z[T[\SU(N)]](m,\zeta)\,
Z[T[\SU(N)]](\zeta,m') 
=\frac{\delta_W(m,m')}{\Delta^2(m)}\ , \label{eq:SS}
\end{equation}
where $\delta_W(m,m')$ is the Weyl-invariant delta function defined
by
\begin{equation}
\delta_W(m,m')= \frac{1}{N!}\sum_{w\in W} (-1)^w \delta(m-w(m'))\ .
\end{equation}
The theory $T[\SU(N)]$ comes from a 4d theory on a segment with an S-duality wall in between two Dirichlet conditions. 
Then it is natural to interpret  $Z[T[\SU(N)]](\zeta,m)$ as the matrix element of
the operator $\cS$ of the 1d quantum mechanical system:
\begin{align}
Z[T[\SU(N)]](\zeta,m)=\langle \zeta | \cS |m \rangle \ ,
\end{align}
where $|m\rangle$ and $|\zeta\rangle$ are two states in $\cH$, the space of functions on  the Cartan of the Lie algebra of the  gauge group $\SU(N)$. 
We assume that the states $|m\rangle$ satisfy
\begin{equation}
\langle m | m' \rangle =\frac{\delta_W(m,m')}{\Delta^2(m)}, \quad\text{and}\quad
\int\! dm \, \Delta^2(m) |m\rangle \langle m| =1\ .
\label{completeset}
\end{equation} 
Then the relation \eqref{eq:SS} is equivalent to $\cS^2=1$.
Intriguingly, the measure factor is  the Vandermonde determinant of $\SL(N,\bR)$ group, not of $\SU(N)$ group we started with.
The wavefunction can also be thought of that of $N$ fermionic particles  moving on a line.
The Hamiltonian cannot, however, be determined from this analysis.

The partition function $Z[T_{\rho}^{\sigma}[\SU(N)]]$ can now be written as 
\begin{equation}
Z[T_{\rho}^{\sigma}[\SU(N)]](\zeta,m)=\langle  \rho,\zeta | \cS | \sigma, m\rangle \ ,
\label{ZTrhosigma}
\end{equation}
where the state $\ket{\rho,\zeta}$ is given by the wavefunction \begin{equation}
\langle m|\rho,\zeta\rangle=\frac{\delta_W(m,\zeta_\rho)}{\Delta(m)\Delta_\rho(\zeta)}\ .
\end{equation}

\section{$T_N$ and $q$-deformed 2d Yang-Mills}\label{sec:comparison}

\begin{figure}
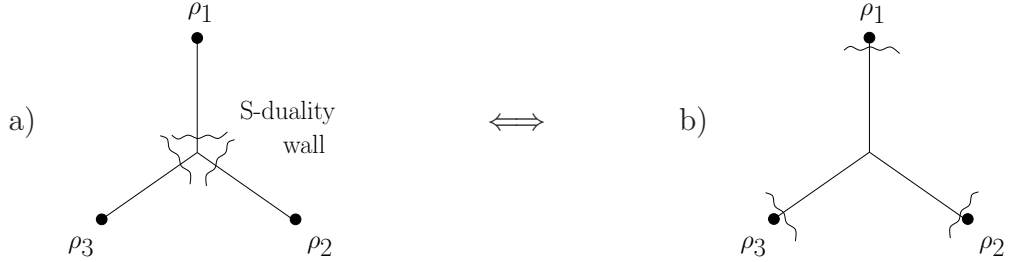

\begin{center}
\[
\text{a)}\quad \vcenter{\hbox{\scalebox{0.4}{\input{graph.pstex_t}}}}
\qquad \qquad \Longleftrightarrow \qquad \qquad
\text{b)}\quad \vcenter{\hbox{\scalebox{0.4}{\input{graph_Sdual.pstex_t}}}}
\]
\end{center}
\caption{4d SYM put on a junction, in two S-dual descriptions. 
a) When three S-duality walls are close to the junction, we have the $T_N$ theory at the center, and Nahm poles $\rho_{1,2,3}$ at the other ends of the segments. 
This figure represents the same theory as that of  Fig.~\ref{fig:trivalent}.
b)
When each S-duality wall is close to each end of three segments,  we have a theory given by a star-shaped quiver.
These theories are mirror to each other.
\label{fig:junction}}
\end{figure}

Another important boundary condition \cite{Benini:2010uu} is defined at
the junction of three edges. 
We consider the simplest one, which gauges the diagonal combination of three $\SU(N)$s. 
Now we put one S-duality wall on each segment, and put the boundary conditions $\rho_{1,2,3}{}_Y$ at the other ends. 
The low-energy limit is the 3d version of $T^{\rho_1,\rho_2,\rho_3}_N$,
which can also arise as the $S^1$ compactification of the 4d $T^{\rho_1,\rho_2,\rho_3}_N$ theory \cite{Benini:2010uu}.
If we put all three S-duality walls very close to the junction, as shown in Fig.~\ref{fig:junction}a,
we have Nahm poles at the ends and the theory $T_N=T_N^{[1,\ldots,1],[1,\ldots,1],[1,\ldots,1]}$ at the center, giving a non-Lagrangian theory. 
If we put all three S-duality walls very close to the ends of three segments, as shown in Fig.~\ref{fig:junction}b, we go to a mirror description which is given by a star-shaped quiver.
In this description the evaluation of the partition function is simple, and we obtain 
\begin{align}
Z[T_N^{\rho_1, \rho_2, \rho_3}](\zeta_1, \zeta_2, \zeta_3)&=\int \!dm\,
\Delta^2(m) \prod_{i=1}^3 Z[T_{\rho_i}^{\sym}[\SU(N)]](\zeta_i,m)  \ ,\\
&= \langle\!\langle\!\langle T_N |\, (\ket{\rho_1,\zeta_1} \otimes \ket{\rho_2,\zeta_2} \otimes \ket{\rho_3,\zeta_3}  )  \ ,
\label{ZTN}
\end{align} where $|T_N \rangle\!\rangle\!\rangle$ is a state in
$\cH^3$ and is the wavefunction determined by the $T_N$ theory: \begin{equation}
Z[T_N](\zeta_1,\zeta_2,\zeta_3)=
\langle\!\langle\!\langle T_N |\, (\ket{\zeta_1} \otimes \ket{\zeta_2} \otimes \ket{\zeta_3} )\ .
\end{equation}

This class includes the 3d version of $E_{6,7,8}$ theory; they are respectively
$T^{[1,1,1],[1,1,1],[1,1,1]}_3$,
$T^{[1,1,1,1],[1,1,1,1],[2,2]}_4$ and
$T^{[1,1,1,1,1,1],[2,2,2],[3,3]}_6$.
The mirror of these theories are Lagrangian theories 
whose quiver diagrams are given by the extended Dynkin diagram of $E_n$.
We will prove the invariance of the partition function of these
theories under the Weyl group of $E_{6,7,8}$ in Appendix~\ref{sec:E}.

This result can be compared with the  superconformal index of the 4d version of $T^{\rho_1,\rho_2,\rho_3}_N$ theory proposed in  \cite{Gadde:2011ik}:
\begin{equation}
\mathcal{I}_{\rho_1, \rho_2, \rho_3}(U_1,U_2,U_3)=\mathcal{N}_{\rho_1, \rho_2, \rho_3}(q)  \sum_\lambda \frac{1}{\textrm{dim}_q {R_\lambda}} 
\prod_{s=1}^3
\mathcal{A}_{\rho_s}(U_s) \chi_{R_\lambda}(U_s) \ . \label{eq:index}
\end{equation}
This formula is proven for $N=2$, and there are many checks for $N>2$ with various types of punctures. 
Let us describe the quantities used in the formula. 
\begin{itemize}
\item $q=e^{-\beta}$ measures the ratio of the radii of $S^1$ and $S^3$.
\item $U_s$ ($s=1,2,3$) is a diagonal matrix encoding the chemical potentials of the flavor symmetries \begin{equation}
U_s=\diag(e^{- i \beta (\zeta_{\rho_s})_j}) \ ,
\end{equation} where 
$\zeta_{\rho_s}$ is chosen
to have the form \eqref{eq:zeta} for $\rho=\rho_s$, see  \cite{Gadde:2011ik}.
\item $R_\lambda$ is the irreducible representation of $\SU(N)$ with highest weight $\lambda$.
\item   $\chi_{R_\lambda}(U)$ is the character of $U=\diag(q^{i\zeta_j})$ in the representation $R_\lambda$, given by the Weyl character formula \begin{equation}
\chi_{R_\lambda}(U)=\frac{\sum_{w\in W} (-1)^w e^{-i \beta \zeta\cdot w(\lambda+\varrho)}  }{i^{\frac{N(N-1)}{2}}\prod_{j<k} \sin \frac{\beta}{2}(\zeta_j-\zeta_k)  } \ ,
\end{equation} where $\varrho$ is the Weyl vector.
\item $\dim_q R_\lambda$ is the $q$-deformed dimension :\begin{equation}
\dim_q R_\lambda= \prod_{j<k}\frac{\sinh\frac{\beta}{2}(\lambda_j-\lambda_k+k-j)}{\sinh \frac{\beta}{2} (k-j) }\ .
\end{equation}
\item $\cA_{\rho}(U)$ is a normalization factor for each $\rho$, known to be \begin{equation}\label{Asym}
\cA_{[1,\ldots,1]}(U)=\exp\left[\sum_{n=1}^\infty\frac{q^n}{1-q^n}\frac1n\chi_\text{adj}(U^n) \right] \ ,
\end{equation} for $\rho=[1,\ldots,1]$;
\begin{equation}\label{A[N-1,1]}
\cA_{[N-1,1]}(U)=\exp\left[\sum_{n=1}^\infty\frac{q^{\frac N2n}}{1-q^n}\frac{a^{Nn}+a^{-Nn}}n\right] \ ,
\end{equation} for $\rho=[N-1,1]$ and  $U=\diag(aq^{\frac{N-2}2},\ldots,aq^{\frac{2-N}2},a^{1-N}) $; 
and \begin{equation}\label{A_[2,2]}
\cA_{[2,2]}(U)=\exp\left[\sum_{n=1}^\infty\frac{q^{n}(1+q^n)}{1-q^n}\frac{a^{2n}+a^{-2n}}n\right] \ ,
\end{equation} for $\rho=[2,2]$  and $U=\diag(aq^{1/2},aq^{-1/2},a^{-1}q^{1/2},a^{-1}q^{-1/2})$.
The general form is not yet known.
\item Finally, $\cN_{\rho_1,\rho_2,\rho_3}$ is a normalization factor independent of the chemical potentials $U$.
\end{itemize}
The $\beta\to 0$ limit  \cite{Dolan:2011rp,Gadde:2011ia,Imamura:2011uw} can be taken nicely by defining \begin{equation}
2\pi m=-\beta \lambda \ ,
\end{equation} keeping $m$ fixed.  This converts the sum over $\lambda$ to an integral over $m$. As for the integrand,
we find\begin{align}
\chi_R(U) &\to  \frac{ \sum_{w\in W} (-1)^w e^{2\pi i m \cdot w(\zeta) } }{ \prod_{j<k} (\zeta_j-\zeta_k) } \ , &
\dim_q R &\to \frac{ \Delta(m) }{ \prod_{j<k} (j-k)  }  \ ,
\end{align} in the $\beta\to 0$ limit, up to a divergent overall factor only depending on $\beta$.
Similarly, 
\begin{equation}
\cA_{\rho}(U)  \to \frac{ \prod_{j<k} ((\zeta_{\rho})_j-(\zeta_{\rho})_k)  } { \Delta_\rho(\zeta) }  \ , \label{eq:aho}
\end{equation} for the known cases $\rho=[1,\ldots,1]$, $\rho=[N-1,1]$ and $\rho=[2,2]$.
See Appendix \ref{ap:qYM_limit} for details.
Assuming the validity of \eqref{eq:aho} for general $\rho$,
the $\beta\to 0$ limit of \eqref{eq:index} becomes \eqref{ZTN}.
This analysis can be thought of as another check of the general proposal in \cite{Gadde:2011ik}, and also a clue to find a general formula for the normalization factor $\cA_{\rho}(U)$. 

Intriguingly, a discrete label $\lambda$ for the representation $R$ becomes a continuous
parameter $m$. The parameter $U$ in the  group became a parameter in the Lie algebra $\zeta$.
 Then the pairing $\chi_R(U)$ became $\vev{m|\cS|\zeta}$, in which 
 $\zeta$ and $m$ play symmetric roles, which was not the case in 4d.

\section{Remarks}\label{sec:conclusions}
In this paper, we studied the partition functions of 3d theory $T^\sigma_\rho[\SU(N)]$ and 
$T^{\rho_1,\rho_2,\rho_3}_N$, which arise as the low-energy limit of 
4d $\cN=4$ SYM placed on a segment and a junction with half-BPS boundary conditions. 
The theory $T^\rho_\sigma[\SU(N)]$ has a linear quiver description, which makes it possible to perform the evaluation of the partition function.  
We saw that the partition function is given by the overlap of wavefunctions given by two boundary conditions, making the 3d mirror symmetry of $T^\rho_\sigma$ and $T^\sigma_\rho$ manifest. 

The theory $T_N^{\rho_1,\rho_2,\rho_3}$ is usually non-Lagrangian, but its 3d mirror has a  Lagrangian description, with which the partition function can be readily evaluated.
We saw that the result again admits an interpretation in terms of the overlap of the wavefunctions.
We also successfully compared  the partition function of $T_N^{\rho_1,\rho_2,\rho_3}$  with the zero-radius limit of the superconformal index of the corresponding 4d theory.
We pointed out that the representation label and the holonomy label both reduce to the labels in the Cartan of the Lie algebra and to play symmetric roles.

One shortcoming in our analysis is that we only studied the 4d theory on a segment in the limit where the length of the segment can be ignored. Therefore we could only identify the wavefunctions of our quantum mechanical system. It would be desirable to study the theory at nonzero length of the segment. This would enable us to determine the Hamiltonian of the 1d theory. 

Another is our cursory analysis of the relation to the $q$-deformed Yang-Mills. 
In \cite{Gadde:2011ik}
the equivalence of the superconformal index with the partition function of the
 $q$-deformed 2d Yang-Mills was discussed. 
The system treated in this note is a reduction on $S^1\times \tilde S^1$ of this higher dimensional system.
This point deserves further study. 
The relation to 3d $SL(N,\mathbb{R})$ Chern-Simons theory \cite{Dimofte:2010tz,Terashima:2011qi} which should be behind the $q$-deformed Yang-Mills will also be interesting.

\section*{Acknowledgments}
The authors thank F.~Benini, T.~Hartman, C.~Herzog, I.~Klebanov and S.~Pufu for discussion.
The work of TN was supported in part by NSF grants PHY-0844827 and PHY-0756966.
The work of YT is supported in part by NSF grant PHY-0969448 and by the Marvin L. Goldberger membership through the Institute for Advanced Study.
He was also supported in part by World Premier International Research Center Initiative (WPI Initiative),  MEXT, Japan through the Institute for the Physics and Mathematics of the Universe, the University of Tokyo.

\appendix 

\section{Examples}\label{sec:example}
In this Appendix we present several explicit examples of our 3d partition functions. These examples serve as the checks of mirror symmetry as well as the expressions given in the main text. 
\subsection{$T_{\rho}^{\sigma}[\SU(4)]$}
\begin{figure}[htbp]
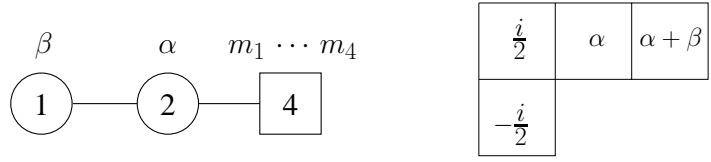

\[
	\vcenter{\hbox{\scalebox{0.45}{\input{1-2-4.pstex_t}}}}\qquad\qquad\quad
	\vcenter{\hbox{\scalebox{0.4}{\input{YoungN4.pstex_t}}}}
\]
	\caption{$T_{[2,1,1]}^{\sym}[\SU(4)]$ theory and  the assignment of the FI parameters to a given Young diagram 	\label{fig:1-2-4}.}
\end{figure}

We begin with $T_{[2,1,1]}^{\sym}[\SU(4)]$ theory whose quiver is given
in Fig.~\ref{fig:1-2-4}.
The partition function of this theory is given by ($\sigma_2=(\sigma_{2,1},\sigma_{2,2})$)
\begin{align}
Z[T_{[2,1,1]}^{\sym}[\SU(4)]]=\int d\sigma_2 d\sigma_1\, \Delta^2(\sigma_2)
 \frac{e^{2\pi i(\alpha[\sigma_2] + \beta\sigma_1)}}{c(\sigma_2-\sigma_1) c(m-\sigma_2)} \ ,
\end{align}
where we used a shorthanded notation
\begin{equation}
d\sigma:=\prod_i d\sigma_i\ , \quad [\sigma]:=\sum_i \sigma_i\ ,
\end{equation}
and
\begin{align}
        	c(\sigma-\rho) := \prod_{i,j}\cosh \pi (\sigma_i - \rho_j) \ , \qquad s(\sigma-\rho) := \prod_{i,j}\sinh \pi (\sigma_i - \rho_j) \
,
\end{align}
for vectors $\sigma = (\sigma_1, \dots , \sigma_k)$ and $\rho = (\rho_1, \dots , \rho_k)$.
After evaluating the integral, we have
\begin{align}
Z[T_{[2,1,1]}^{\sym}[\SU(4)]]=\frac{1}{\Delta(m)\tilde{\Delta}(\alpha)}\left[ e^{2\pi i
(\alpha+\beta) m_1 +\alpha m_2} \sinh \pi (m_3-m_4) \pm
 (\textrm{perm.})\right] \ ,
\label{Z421}
\end{align}
where we have 4! signed permutations of $m_i$ such that the expression
is invariant under permutations of $m_i$, and we have defined
\begin{align}
\tilde{\Delta}(\alpha)=\cosh(\pi \alpha) \sinh(\pi \beta) \cosh(\pi(\alpha+\beta)) \ .
\end{align}
The first term inside the bracket can be written as
\begin{equation}
\sum_{w\in W} e^{2\pi i {\zeta}\cdot w(m)} \qquad\quad  \text{where}\quad 
{\zeta}=\left(\alpha+\beta, \alpha, \frac{i}{2}, -\frac{i}{2}\right).
\end{equation}
Moreover, $\tilde{\Delta}(\alpha)$ is equivalent to $\Delta_{[2,1,1]}(\zeta)$.
The graphical rule for representing ${\zeta}$ is also shown in Fig.~\ref{fig:1-2-4}.

\begin{figure}[htbp]
	\begin{center}
	\scalebox{0.5}{\input{1-2-2.pstex_t}}
	\caption{A quiver diagram of $T^{[2,1,1]}_{\sym}[\SU(4)]$ theory.}
	\label{fig:1-2-2}
	\end{center}
	\end{figure}

A mirror theory is a $T^{[2,1,1]}_{\sym}[\SU(4)]$ theory with a quiver shown in Fig~\ref{fig:1-2-2}.
 The partition function is given by
\begin{multline}\label{Z1-2-2}
	Z[T^{[2,1,1]}_{\sym}[\SU(4)]] =\int\! d\sigma_3d\sigma_2 d\sigma_1\, \Delta^2(\sigma_2) \Delta^2(\sigma_3) \\
	 \cdot \frac{e^{2\pi i(\alpha[\sigma_3] + \beta[\sigma_2] + \gamma \sigma_1)}}{c(\sigma_2-\sigma_1) c(m_1-\sigma_2) c(\sigma_3-\sigma_2) c(\sigma_3- m_2 ) c(\sigma_3- m_3 )} \ ,
\end{multline} 
where we used the shorthanded notation: $\sigma_i = (\sigma_{i,1}, \sigma_{i,2}) ~(i=2,3)$.
After integration, one finds that it takes the same form as in \eqref{Z421} with the change of 
the parameters:
\begin{align}
	&m_1 \to -\alpha-\frac{\beta}{2} \ , \qquad m_2 \to -\frac{\beta}{2} \ , \qquad m_3 \to \frac{\beta}{2} \ , \\
	&\alpha \to m_3 - m_4 \ , \qquad \beta \to m_2 - m_3 \ , \qquad \gamma \to m_1 - m_2 \ , \nonumber
\end{align}
and we thus checked the mirror symmetry between the $T^{[2,1,1]}_{\sym}[\SU(4)]$ and $T^{\sym}_{[2,1,1]}[\SU(4)]$
theories by explicit calculation of the partition functions.

\bigskip

\begin{figure}[htbp]
	\begin{center}
	\scalebox{0.5}{\input{2-1-1-1.pstex_t}}
	\caption{A quiver diagram of $T^{[2,1,1]}_{[2,1,1]}[\SU(4)]$ theory.}
	\label{fig:2-1-1-1}
	\end{center}
	\end{figure}

We give one more example with self-mirror symmetry: $T^{[2,1,1]}_{[2,1,1]}[\SU(4)]$ theory,
with the quiver diagram shown in Fig.~\ref{fig:2-1-1-1}.
The partition function is given by
\begin{align}\label{Z2-1-1-1}
	Z[&T^{[2,1,1]}_{[2,1,1]}[\SU(4)]] = \int d\sigma_1 d\sigma_2 \frac{e^{2\pi i(\alpha \sigma_2 + \beta\sigma_1)}}{
	\cosh\pi(\sigma_2-m_1)\cosh\pi(\sigma_2-m_2)\cosh\pi(\sigma_2-\sigma_1) \cosh\pi(\sigma_1-m_3)} \nonumber\\
	&= -i \Bigg[ \frac{e^{2\pi i(\alpha + \beta)m_1}}{\sinh\pi m_{12}\cosh\pi m_{13} \sinh\pi\beta\cosh\pi(\alpha+\beta)}
	-  \frac{e^{2\pi i(\alpha + \beta)m_2}}{\sinh\pi m_{12}\cosh\pi m_{23} \sinh\pi\beta\cosh\pi(\alpha+\beta)}\nonumber\\
	& +  \frac{e^{2\pi i(\alpha m_3 + \beta m_1)}}{\sinh\pi m_{12}\cosh\pi m_{13} \cosh\pi\alpha\sinh\pi\beta}
	-  \frac{e^{2\pi i(\alpha m_3 + \beta m_2)}}{\sinh\pi m_{12}\cosh\pi m_{23} \cosh\pi\alpha\sinh\pi\beta}\nonumber\\
	& +  \frac{i e^{2\pi i(\alpha + \beta) m_3}}{\cosh\pi m_{13}\cosh\pi m_{23} \cosh\pi\alpha\cosh\pi(\alpha+\beta)} \Bigg] \ ,
\end{align}
where $m_{ij} := m_i - m_j$.
This can be put into the form \eqref{eq:general}.
This partition function is invariant under the following replacements of the parameters:
\begin{align}
	m_1 \leftrightarrow \frac{\alpha + \beta}{2} \ , \qquad m_2 \leftrightarrow  \frac{\alpha - \beta}{2} \ , 
	\qquad m_3 \leftrightarrow -\frac{\alpha + \beta}{2} \ .
\end{align}
We have confirmed that the $T^{[2,1,1]}_{[2,1,1]}[\SU(4)]$ theory is self-mirror as is expected.

\subsection{$\SU(M)$ Theory with $N$ Flavors}\label{ap:M-N}
\begin{figure}[htbp]
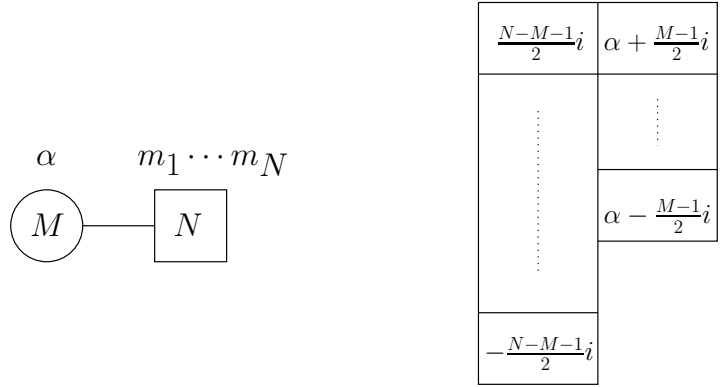

	\begin{center}\[
	\vcenter{\hbox{\scalebox{0.5}{\input{M-N.pstex_t}}}}
	\qquad \qquad \qquad\qquad
	\vcenter{\hbox{\scalebox{0.5}{\input{M-N-Young.pstex_t}}}}
	\]
	\caption{$\SU(M)$ theory with $N$ flavors. A corresponding Young diagram is also shown with
	an assignment of the FI parameters.}
	\label{fig:M-N}
	\end{center}
\end{figure}
Next, let us consider $T_{[N-M,M]}[\SU(N)]$ theory, namely, 
$\SU(M)$ theory with $N$ flavors. The corresponding quiver and the Young diagram are shown in Fig.~\ref{fig:M-N}. The partition function  is given by 
\begin{align}
	Z[T_{[N-M,M]}[\SU(N)]] = \int \!d\sigma \, \Delta^2 (\sigma) \frac{e^{2\pi i \alpha \sum_{i=1}^M \sigma_i}}{\prod_{a=1}^N\prod_{i=1}^M\cosh\pi(\sigma_i - m_a)} \ . \label{eq:NM}
\end{align}
One can integrate out $\sigma_i$ one by one picking up a single pole at $m_a + \frac{2n+1}{2}i~ (n=0,1,\dots)$.
Each pole contributes a term $(-1)^{nN} e^{-(2n+1)\pi\alpha}$
that can be summed up to be 
\begin{align}
	&\frac{1}{\sinh\pi\alpha} \qquad\quad \text{for $N$: even} \ , &
	&\frac{1}{\cosh\pi\alpha} \qquad\quad \text{for $N$: odd}  \ .
\end{align}
Then we obtain 
\begin{multline}\label{eq:ZNM}
	Z[T_{[N-M,M]}[\SU(N)]] 
	 =\frac{1}{\tilde{\Delta}(\alpha)\Delta (m)}\Bigl[ e^{2\pi i \alpha \sum_{i=1}^M m_i} \prod_{i<j\le M}
	\sinh\pi (m_i - m_j) \\
	\cdot \prod_{M<a<b}\sinh\pi (m_a - m_b) \pm \text{(perm)}\Bigr] \ ,
\end{multline}
where 
\begin{align}
	\tilde{\Delta}(\alpha) =
\begin{cases}
  ~(\sinh\pi\alpha)^{M} \qquad & \text{for $N$: even} \ , \nonumber\\ 
	~ (\cosh\pi\alpha)^{M} \qquad & \text{for $N$: odd} \ . \nonumber
\end{cases}
\end{align}
This is reproduced by \eqref{eq:general}.

\section{Inductive Proof of $Z[T_{\rho}[\SU(N)]]$}\label{sec:proof}

Here we prove the expression for $Z[T_{\rho}^{\sigma}[\SU(N)]]$ \eqref{eq:general} when $\sigma=[1,\ldots,1]$. The proof
proceeds by induction with respect to the length of the quiver, and
computation is essentially the same as in the previous example in
Appendix \ref{ap:M-N}.

Suppose that \eqref{eq:general} holds for $T_\rho[\SU(N_M)]$, whose quiver is given by
\begin{align}
\SU(N_M)-\U(N_{M-1})-.....-\U(N_1)\ .
\end{align}
We would like to show that \eqref{eq:general} is true for $T_{\rho'}[\SU(N_{M+1})]$,
which has a quiver
\begin{align}
\SU(N_{M+1})-\U(N_M)-\U(N_{M-1})-.....-\U(N_1)\ .
\end{align}

We begin with the expression for $Z[T_{\rho}[\U(N_M)]]$ (we neglect the
overall normalization constant in this Appendix):
\begin{align}
Z[T_{\rho}(\U(N_{M}))](\sigma_M, \zeta)=
\frac{\sum_{w\in \mathfrak{S}_{N_{M}}} (-1)^w e^{2\pi i \sigma_M\cdot w(\zeta_{\rho})}}{\Delta_{\rho}(\zeta)
 \Delta(\sigma_M)} \ ,
\end{align}
where we do not impose the constraint $\sum_i \sigma_{M,i}=0$.
We then have
\begin{align*}
Z[T_{\rho'}[\SU(N_{M+1})]]& =\int\! d\sigma_M\, \Delta^2(\sigma_M)
 \frac{1}{c(\sigma_M-m_{M+1})}Z[T_{\rho}[\U(N_{M})]](\sigma_M,
 \zeta_{\rho}) \ ,\\
&=   
 \frac{1}{\Delta_{\rho}(\zeta)}
 \int\! d\sigma_M\, \Delta(\sigma_M) \frac{\sum_{w\in
 \mathfrak{S}_{N_{M}}}(-1)^w e^{2\pi i \sigma_M \cdot w(\zeta_{\rho})}}{c(\sigma_M-m_{M+1})}
\ ,
\end{align*}
where we used the notation of Appendix \ref{sec:example}.
We can evaluate this integral by closing the contour and summing up
contributions from poles in the integrand. For an integral with respect
to $\sigma_{M,i}$, the poles are located at
$\sigma_{M,i}=m_{4,k}+i(n+1/2)$, where $k \in J:=\{ 1, 2,\ldots,
N_{M+1}\}$ and $n\ge 0$. 
Due to the existence of
$\Delta(\sigma_M)$ factor in the integrand (which vanishes whenever
two $\sigma_{M,i}$'s coincide), we need to choose different
$k$'s for different $i$'s. Let us denote this by $k\in I$, where
$I$ is a subset of $J$ such that $|I|=N_M$.
The summation over an integer $n$ gives $1/\sinh \pi (\zeta_{\rho})_{w(i)}$ or
$1/\cosh \pi (\zeta_{\rho})_{w(i)}$ depending on
whether $N_M$ is even or odd. 
After evaluating the integrals we have
\begin{align}
Z[T_{\rho'}[\SU(N_{M+1})]]=
\frac{ 
 \displaystyle\sum_I 
 \displaystyle\sum_{w\in \mathfrak{S}_{N_M}}(-1)^w (-1)^I 
 \, e^{2\pi i \sum_{j\in I} m_{M,j} (\zeta_{\rho})_{w(j)}}
 \displaystyle \prod_{i<j, i,j\in J\backslash I} \sinh \pi (m_i-m_j)
}
{ 
  \Delta(m)
  \tilde{\Delta}(\zeta)
} \ ,
\end{align}
where $(-1)^I$ is a sign defined by
\begin{equation*}
(-1)^I=(-1)^{\# \{ (i,j) | i\in I, j\in J\backslash I, i>j\}} \ ,
\end{equation*}
 and 
\begin{align*}
\tilde{\Delta}(\zeta):=
\begin{cases}\Delta_{\rho}(\zeta) \ \prod_{i \in
 I} \sinh \pi (\zeta_{\rho})_{w(i)}\qquad\quad \textrm{for ~$N_M$: even} \ , \\
\Delta_{\rho}(\zeta) \, \prod_{i \in
 I} \cosh \pi (\zeta_{\rho})_{w(i)}\qquad\quad  \textrm{for ~$N_M$: odd}
\ .
\end{cases}
\end{align*}
Now by expanding $\sinh$ into a sum of two exponentials, we have
\begin{align*}
 \displaystyle \prod_{i<j,\  i,j\in J\backslash I} \sinh \pi (m_i-m_j)
  = \sum_{w^* \in \mathfrak{S}_{(N_{M+1}-N_{M})}} (-1)^{w^*}e^{2\pi \zeta^* \cdot
 w^*(m^*)} \ ,
\end{align*}
where $m^*=(m_{i})_{i\in J\backslash I}$ and
$\zeta^*=(\frac{N_{M+1}-N_M-1}{2} i, \ldots, -\frac{N_{M+1}-N_M-1}{2}i)$.
We therefore have
\begin{align*}
Z[T_{\rho'}[\SU(N_{M+1})]]=
\frac{1}
{ 
  \Delta(m)
  \tilde{\Delta}(\zeta)
}
 &\sum_I \sum_{w\in \mathfrak{S}_{N_M}}
  \sum_{w^*\in \mathfrak{S}_{(N_{M+1}-N_M)}}(-1)^w (-1)^{w^*} (-1)^I \\
& \times e^{2\pi i (\sum_{j\in I} m_{M,j} (\zeta_{\sigma})_{w(j)} + \sum_{j\in
 J\backslash I} m^*_{j} \zeta^*_{w^*(j)})}
\ .
\end{align*}
The sums over $w, w^*$ and $I$ can be rewritten as a single sum over $w'\in
\mathfrak{S}_{N_M+1}$, when we define $\zeta'=(\zeta_{\rho}, \zeta^*)$
\begin{align}
Z[T_{\rho'}[\SU(N_{M+1})]]=
\frac{ 
 \sum_{w' \in \mathfrak{S}_{N_{M+1}}}(-1)^{w'}
 e^{2\pi i (\sum_{j\in J} m_{M,j} (\zeta')_{w(j)})}
}
{ 
  \Delta(m)
  \tilde{\Delta}(\zeta) 
}
\ .
\end{align}
We can verify that $\zeta'$ coincides with $\zeta_{\rho'}$, and
$\tilde{\Delta}(\zeta)$ with $\Delta_{\rho'}(\zeta)$. This is what we
wanted to show.

\section{A Consistency Condition for $T_{\rho}^{\sigma}[\SU(N)]$ }\label{sec:dominant}

In this Appendix we prove that $Z[T_{\rho}^{\sigma}[\SU(N)]]$ vanishes
when $T_{\rho}^{\sigma}[\SU(N)]$ is empty, that is, when the
condition $\sigma^T\ge \rho$ is not satisfied.

In the expression for the partition function given in \eqref{eq:general}, we
need to compute a pairing of two $N$-vectors $\zeta_{\sigma}$ and a
permutation of $\zeta'_{\rho}$. Consider $(i,a)$ and $(j,a)$ in the same
column $a$ of $\sigma$ and another two boxes $(k,b)$ and $(l,b)$ in the
same column $b$ of $\rho$. Then we can consider two possible pairing of
these four boxes: (1) $(i,a)$ with $(k,b)$ and $(j,a)$ with $(l,b)$
and (2) $(i,a)$ with $(l,b)$ and $(j,a)$ with $(k,b)$. These parings are
related by a single permutation, and the corresponding contributions
cancel out in the expression \eqref{eq:general} due to the following
identity
\begin{equation}
e^{2\pi i [\zeta_{(i,a)} \zeta'_{(k,b)}+ \zeta_{(j,a)}\zeta'_{(l,b)} ]}
-
e^{2\pi i [\zeta_{(i,a)} \zeta'_{(l,b)}+ \zeta_{(j,a)}\zeta'_{(k,b)} ]}
=0 \ .
\end{equation}
This is shown by noting that $\zeta_{(i,a)}-\zeta_{(j,a)}\in \mathbb{Z}$ and
$\zeta'_{(k,b)}-\zeta'_{(l,b)} \in \mathbb{Z}$.

This means that non-zero contributions are possible only when any
two boxes in the same column of $\rho$ pair up with boxes in different
columns of $\sigma$. For example, let us denote the number of boxes in
the $i$-th column of $\rho$ ($\sigma^T$) by $n_i$ ($m_i$).
Now $n_1$ boxes in the first
column of $\rho$ should all pair up with
boxes in different columns of $\sigma$. This means 
$\sigma$
should have at least $n_1$ columns, namely $m_1\ge n_1$. Similarly, $n_1+n_2$ boxes in the first
and second columns of $\rho$ should be distributed to $\sigma$ in such a way that at most two boxes
are in the same column of $\sigma$. This means $m_1+m_2\ge n_1+n_2$.
By repeating this we conclude $\sigma^T\ge \rho$.

\section{The $\beta\to 0$ Limit of $q$-deformed Yang-Mills}\label{ap:qYM_limit}
We here check \eqref{eq:aho} for the known examples 
$\rho=[1,\dots,1]$, $\rho=[N-1,1]$ and $\rho=[2,2]$.

Firstly, the square of the normalization factor $\cA_{[1,\dots,1]}$ is an
inverse of the vector multiplet contribution to the index.
The $\beta\to 0$ limit is given in (3.7) of \cite{Gadde:2011ia} and it takes the form of \eqref{eq:aho}.

Secondly, \eqref{A[N-1,1]} can be written as 
\begin{align}
	\cA_{[N-1,1]} &= \prod_{m=0}^\infty \frac{1}{(1-a^N q^{\frac{N}{2}+m})(1-a^{-N} q^{\frac{N}{2}+m})} \nonumber\\
	&\stackrel{\rm reg}{=} \prod_{m=0}^\infty \frac{[m+\frac{N}{2}]_q}{[m+\frac{N}{2}+i \zeta]_q} 
	\frac{[m+\frac{N}{2}]_q}{[m+\frac{N}{2}-i \zeta]_q}  \ ,
\end{align}
where $[n]_q := \frac{1-q^n}{1-q}$ is the $q$-integer, and we have set $a^N = q^{i\zeta}$.
In the last line, we introduced a regularization to subtract a
divergence in a manner that does not depend on $\zeta$ .
Then one can take $\beta\to 0$ ($q\to 1$) limit and obtain
\begin{align}\label{ap:A_[N-1,1]}
	\cA_{[N-1,1]} &= \prod_{m=0}^\infty \left( 1+ \frac{\zeta^2}{(m+\frac{N}{2})^2}  \right)^{-1} =
	\begin{cases}
	\displaystyle
		\frac{\pi \zeta}{\sinh\pi\zeta} \prod_{m=1}^{\frac{N}{2}-1}(\zeta^2 + m^2) \qquad\qquad\qquad \text{for $N:$ even} \ , \\
		\displaystyle
		\frac{1}{\cosh\pi\zeta} \prod_{m=1}^{\frac{N-1}{2}}\left(\zeta^2 + (m-\frac{1}{2})^2\right) \qquad\, \text{for $N:$ odd} \ .
\end{cases}
\end{align}
Recall that $U= \diag (aq^{\frac{N-2}{2}},\dots,aq^{\frac{2-N}{2}},a^{1-N})$ and equivalently
\begin{align}
	\zeta_{[N-1,1]} = \left(\frac{\zeta}{N} + \frac{N-2}{2}i, \dots, \frac{\zeta}{N} - \frac{N-2}{2}i, \frac{1-N}{N}\zeta\right) \ ,
\end{align}
and it follows that \eqref{ap:A_[N-1,1]} can be written as \eqref{eq:aho}.

Finally, \eqref{A_[2,2]} becomes
\begin{align}
	\cA_{[2,2]} &\stackrel{\rm reg}{=}  \prod_{m=0}^\infty \frac{[m+1]_q}{[m+1+i \zeta]_q} 
	\frac{[m+1]_q}{[m+1-i \zeta]_q} \frac{[m+2]_q}{[m+2+i \zeta]_q} 
	\frac{[m+2]_q}{[m+2-i \zeta]_q}  \nonumber\\
	&\stackrel{\beta\to 0}{\rightarrow}  
	\left( \frac{\pi \zeta}{\sinh\pi\zeta}\right)^2(\zeta^2 + 1) \ .
\end{align}
Here we set $a^2=q^{i\zeta}$ and the above result is the same form
as \eqref{eq:aho} with
\begin{align}
	\zeta_{[2,2]} = \left( \frac{\zeta-i}{2}, \frac{\zeta+i}{2}, 
	-\frac{\zeta +i}{2}, -\frac{\zeta -i}{2}\right) \ .
\end{align}

\section{$E_n$ Invariance of the Partition Function}\label{sec:E}
Here we show that the partition function of 3d $\cN=4$ quiver theory whose form is given by the $E_n$ Dynkin diagram  has $E_n$ symmetry. 
As a preparation, let us consider $T_{[N,N]}(\SU(2N))$ theory, 
whose partition function is given in \eqref{eq:ZNM}. 
Let us split the flavor symmetry $\U(2N)$ into $\U(n_1)\times \cdots \times \U(n_k)$ with $\sum n_j=2N$,
and introduce FI and mass parameters $\beta_{j}, m_{j,a_j}$ $(j=1,\dots,k, a_j=1,\dots,n_j)$ which correspond to a factor $e^{2\pi i \sum_j  \beta_j [m_j] }$ with $[m_j] = \sum_{a=1}^{n_j} m_{j,a}$. Then the partition function is invariant under the operation \begin{equation}
(\alpha,\beta_j) \mapsto  (-\alpha,\beta_j+\alpha) \ .\label{app:weylaction}
\end{equation}
This represents the Weyl reflection acting on the FI parameters:
let us introduce vectors $\vec v$ and $\vec w_j$ satisfying $(\vec
v,\vec v)=2$, $(\vec w_j,\vec w_{j'})=2\delta_{j j'}$ and $(\vec v,\vec
w_j)=-1$. We also package the FI parameters into a vector
$\vec{\alpha}$ satisfying $\alpha = (\vec v,\vec\alpha)$ and $\beta_i = (\vec w_i,\vec\alpha)$.
Then the action \eqref{app:weylaction} is the Weyl reflection with respect to $\vec v$, \begin{equation}
\vec \alpha \to R_{\vec{v}}\vec\alpha =\vec \alpha - (\vec \alpha,\vec v)\vec v \ .
\end{equation}

\begin{figure}[htbp]\[
 \quad \vcenter{\hbox{\scalebox{0.45}{\input{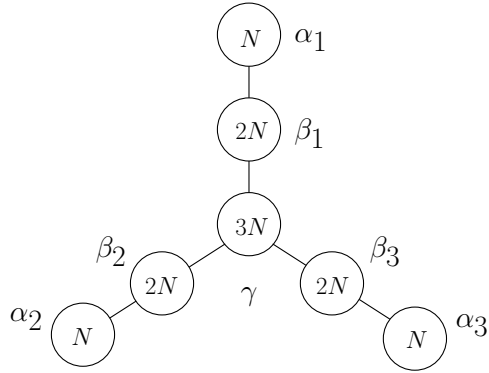}}}}
\]
 \caption{Quiver of the rank-$N$ version of the $E_6$ theory. \label{fig:app}}
\end{figure}

Therefore, for the $E_6$ quiver shown in
Fig.~\ref{fig:app}, the partition function has the invariance under
the Weyl reflections which act
\begin{subequations}
\begin{align}
(\alpha_i, \beta_i)  &\to (-\alpha_i, \alpha + \beta_i)   \quad \quad&&       \text{for a specific $i$}, \\
(\alpha_i, \beta_i, \gamma)   &\to (\alpha_i + \beta_i, -\beta ,
 \gamma+\beta_i)  &&  \text{for a specific $i$, and} \\
(\gamma, \beta_1,\beta_2,\beta_3) &\to (-\gamma, \beta_1 + \gamma, \beta_2 + \gamma, \beta_3 + \gamma) \ .
\end{align}
\end{subequations}			      
They generate the affine Weyl group of $E_6$ acting on $\alpha_{1,2,3}$, $\beta_{1,2,3}$ and $\gamma$.
The affine shift corresponds to the  FI parameter of the diagonal, completely decoupled $\U(1)$ and does not matter. 
Thus we conclude that the partition function is symmetric under $E_6$, which is not manifest in the UV Lagrangian description. Similarly, we can show the $E_{7,8}$ invariance of the partition function of the quiver of the form of Dynkin diagram of $E_{7,8}$. 

In general, it was argued in \cite{Gaiotto:2008ak} using the property of the monopole operators that the flavor symmetry of the Coulomb branch of a `good' $\SU$ quiver theory is given by the group whose Dynkin diagram is given by the sub-quiver formed by the `balanced' nodes of the quiver. This statement can be proved at the level of the partition function using exactly the same argument as above.

\bibliographystyle{utphys}
\small\baselineskip=.8\baselineskip
\bibliography{paper}

\providecommand{\href}[2]{#2}\begingroup\raggedright\begin{thebibliography}{10}

\bibitem{Gaiotto:2008ak}
D.~Gaiotto and E.~Witten, ``{S-Duality of Boundary Conditions in
  ${\mathcal{N}}\!=4$ Super Yang-Mills Theory},''
\href{http://arxiv.org/abs/0807.3720}{{\ttfamily arXiv:0807.3720 [hep-th]}}.

\bibitem{Kapustin:2009kz}
A.~Kapustin, B.~Willett, and I.~Yaakov, ``{Exact Results for Wilson Loops in
  Superconformal Chern- Simons Theories with Matter},''
  \href{http://dx.doi.org/10.1007/JHEP03(2010)089}{{\em JHEP} {\bfseries 03}
  (2010) 089},
\href{http://arxiv.org/abs/0909.4559}{{\ttfamily arXiv:0909.4559 [hep-th]}}.

\bibitem{Gaiotto:2009we}
D.~Gaiotto, ``{${\mathcal{N}}\!=2$ Dualities},''
\href{http://arxiv.org/abs/0904.2715}{{\ttfamily arXiv:0904.2715 [hep-th]}}.

\bibitem{Gadde:2011ik}
A.~Gadde, L.~Rastelli, S.~S. Razamat, and W.~Yan, ``{The 4d Superconformal
  Index from $q$-deformed 2d Yang- Mills},''
\href{http://arxiv.org/abs/1104.3850}{{\ttfamily arXiv:1104.3850 [hep-th]}}.

\bibitem{Benini:2010uu}
F.~Benini, Y.~Tachikawa, and D.~Xie, ``{Mirrors of 3d Sicilian Theories},''
  \href{http://dx.doi.org/10.1007/JHEP09(2010)063}{{\em JHEP} {\bfseries 09}
  (2010) 063},
\href{http://arxiv.org/abs/1007.0992}{{\ttfamily arXiv:1007.0992 [hep-th]}}.

\bibitem{Dolan:2011rp}
F.~A.~H. Dolan, V.~P. Spiridonov, and G.~S. Vartanov, ``{From 4d Superconformal
  Indices to 3d Partition Functions},''
\href{http://arxiv.org/abs/1104.1787}{{\ttfamily arXiv:1104.1787 [hep-th]}}.

\bibitem{Gadde:2011ia}
A.~Gadde and W.~Yan, ``{Reducing the 4d Index to the S$^3$ Partition
  Function},''
\href{http://arxiv.org/abs/1104.2592}{{\ttfamily arXiv:1104.2592 [hep-th]}}.

\bibitem{Imamura:2011uw}
Y.~Imamura, ``{Relation Between the 4d Superconformal Index and the S$^3$
  Partition Function},''
\href{http://arxiv.org/abs/1104.4482}{{\ttfamily arXiv:1104.4482 [hep-th]}}.

\bibitem{Benvenuti:2011ga}
S.~Benvenuti and S.~Pasquetti, ``{3D-Partition Functions on the Sphere: Exact
  Evaluation and Mirror Symmetry},''
\href{http://arxiv.org/abs/1105.2551}{{\ttfamily arXiv:1105.2551 [hep-th]}}.

\bibitem{Gulotta:2011si}
D.~R. Gulotta, C.~P. Herzog, and S.~S. Pufu, ``{From Necklace Quivers to the
  F-theorem, Operator Counting, and T(U(N))},''
\href{http://arxiv.org/abs/1105.2817}{{\ttfamily arXiv:1105.2817 [hep-th]}}.

\bibitem{Nakajima}
H.~Nakajima, ``Instantons on ale spaces, quiver varieties and kac-moody
  algebras,'' \href{http://dx.doi.org/10.1215/S0012-7094-94-07613-8}{{\em Duke
  Math. Journal} {\bfseries 76} (1994) 365}.

\bibitem{Hanany:1996ie}
A.~Hanany and E.~Witten, ``{Type IIB Superstrings, BPS Monopoles, and Three-
  Dimensional Gauge Dynamics},''
  \href{http://dx.doi.org/10.1016/S0550-3213(97)00157-0}{{\em Nucl. Phys.}
  {\bfseries B492} (1997) 152--190},
\href{http://arxiv.org/abs/hep-th/9611230}{{\ttfamily arXiv:hep-th/9611230}}.

\bibitem{Kanno:2009ga}
S.~Kanno, Y.~Matsuo, S.~Shiba, and Y.~Tachikawa, ``{${\mathcal{N}}\!=2$ Gauge
  Theories and Degenerate Fields of Toda Theory},''
  \href{http://dx.doi.org/10.1103/PhysRevD.81.046004}{{\em Phys. Rev.}
  {\bfseries D81} (2010) 046004},
\href{http://arxiv.org/abs/0911.4787}{{\ttfamily arXiv:0911.4787 [hep-th]}}.

\bibitem{Drukker:2010vg}
N.~Drukker and F.~Passerini, ``{(De)Tails of Toda CFT},''
\href{http://arxiv.org/abs/1012.1352}{{\ttfamily arXiv:1012.1352 [hep-th]}}.

\bibitem{Hosomichi:2010vh}
K.~Hosomichi, S.~Lee, and J.~Park, ``{AGT on the S-Duality Wall},''
  \href{http://dx.doi.org/10.1007/JHEP12(2010)079}{{\em JHEP} {\bfseries 12}
  (2010) 079},
\href{http://arxiv.org/abs/1009.0340}{{\ttfamily arXiv:1009.0340 [hep-th]}}.

\bibitem{Kapustin:1999ha}
A.~Kapustin and M.~J. Strassler, ``{On mirror symmetry in three-dimensional
  Abelian gauge theories},'' {\em JHEP} {\bfseries 9904} (1999) 021,
  \href{http://arxiv.org/abs/hep-th/9902033}{{\ttfamily arXiv:hep-th/9902033
  [hep-th]}}.

\bibitem{Kapustin:2010xq}
A.~Kapustin, B.~Willett, and I.~Yaakov, ``{Nonperturbative Tests of
  Three-Dimensional Dualities},''
  \href{http://dx.doi.org/10.1007/JHEP10(2010)013}{{\em JHEP} {\bfseries 1010}
  (2010) 013}, \href{http://arxiv.org/abs/1003.5694}{{\ttfamily arXiv:1003.5694
  [hep-th]}}.

\bibitem{Dimofte:2010tz}
T.~Dimofte, S.~Gukov, and L.~Hollands, ``{Vortex Counting and Lagrangian
  3-Manifolds},''
\href{http://arxiv.org/abs/1006.0977}{{\ttfamily arXiv:1006.0977 [hep-th]}}.

\bibitem{Terashima:2011qi}
Y.~Terashima and M.~Yamazaki, ``{$SL(2,\mathbb{R})$ Chern-Simons, Liouville,
  and Gauge Theory on Duality Walls},''
\href{http://arxiv.org/abs/1103.5748}{{\ttfamily arXiv:1103.5748 [hep-th]}}.

\end{thebibliography}\endgroup

\end{document}